\documentstyle[psfig]{l-aa}
 
\def\te{$T_{\mathrm{eff}}$ }

\begin{document}
 
\thesaurus{emission line, Be: 08.05.2; evolution: 08.05.3; rotation: 08.18.1}

\title{Differences in the fractions of Be stars in galaxies.}


\author{Andr\'e Maeder$^1$ \and  Eva K.\ Grebel$^{2,3,}$\thanks{Hubble Fellow}
\and Jean--Claude Mermilliod$^4$}
\institute{Observatoire de Gen\`eve, CH-1290 Sauverny, Switzerland, 
	 e--mail:~Andre.Maeder@obs.unige.ch \and 
UCO/Lick Observatory, University of California, Santa Cruz, CA 95064, USA
\and University of Washington, Department of Astronomy, Box 351580,
     Seattle, WA 98195, USA
\and Institut d'Astronomie, UniL, CH--1290 Chavannes--des--Bois,
Switzerland}

\date{Received date; accepted date}

\maketitle

\markboth{A. Maeder et al.: Differences of the fractions of Be stars in 
galaxies.}{A. Maeder et al.: Differences of the fractions of Be stars in 
galaxies.}


\begin{abstract}
The number ratios  $Be/(B+Be)$  of Be to B--type stars
 in young, well
studied clusters of the Galaxy, the LMC and SMC are examined. In order to
disentangle age and metallicity effects we choose clusters  
in the same age interval 
and for which reliable photometric and spectroscopic 
data are available. Number counts
are made for various magnitude intervals, and the results are
found to be stable with respect to this choice. In the magnitude interval
$M_{\mathrm{V}} = -5$ to -1.4 (i.e. O9 to B3) we obtained a
 ratio $Be/(B+Be)$ =
0.11, 0.19, 0.23, 0.39 for 21 clusters located
in the interior of the Galaxy, the exterior of the Galaxy, the
LMC and the SMC, respectively.

Various hypotheses for these differences are examined. An interesting
possibility
is that the average rotation is faster at low metallicities as a result
of star formation processes. The much higher relative N--enrichment
found by Venn et al. (\cite{vencar}) in A--type supergiants of the
 SMC, compared to galactic
supergiants, also strongly supports the
presence of more rotational mixing at low metallicities.
We discuss whether high rotational mixing may be
 the source of primary nitrogen 
in the early chemical evolution of galaxies. 
\end{abstract}

\section{Introduction}
 
About  50\% of the B stars in the SMC cluster NGC 330 were
found to be Be stars by Feast (\cite{fea}). This high fraction,
 compared to  10 to 20\% in the
Milky Way, was confirmed by subsequent studies done  by Grebel 
 (\cite{greric, grerob}), Mazzali et al. (\cite{maz}), and Keller et al.
(\cite{kel2}). A pronounced difference between the Be star
content in the Magellanic Cloud clusters and in the Milky Way was also found
by Grebel et al. (\cite{grerobwil}).

The Be phenomenon is closely related to fast rotation.
Suggestions were made in the past that Be stars occur near the end of the main 
sequence (MS) phase; however, extended studies in clusters have shown the
occurrence of Be stars from the zero--age sequence to the end of the MS
(Mermilliod \cite{mer}). Mermilliod has also found that
 the fraction of Be
 stars with respect to normal B stars is low in very young clusters, that it 
reaches a maximum for clusters
 with turnoff in the range of O9 to B3 and that it
declines after that.
 Evidences for an age effect are also well shown by Fig. 5
from Grebel (\cite{gre}).

Be stars are generally slightly shifted to the red of the MS band
(Grebel et al. \cite {grerob}; cf. also Fig. 1 below).
 This effect is consistent with the effect
of rotational reddening found by Maeder and Peytremann (\cite{maepey})
and also with the effect of a circumstellar envelope, which 
contributes to a reddening  as well as to the formation of the emission 
lines (Gehrz et al.\ 1994, Cot\'e \& Waters 1987).
A physical model of the radiation-driven wind in rotating
stars  has been proposed by Bjorkman and Cassinelli (\cite{bjo}).
This model, the so-called wind compressed disk (WCD) model,
predicts that the isotropic wind particles travel along trajectories
that go through the equatorial plane, where shocks occur. 
The gas is thus confined there  and forms a dense equatorial disk.
Owocki et al. (\cite{owoc}) have shown that gravity darkening,
as predicted by the von Zeipel theorem, leads to a very different
wind morphology, where the disk is more difficult to form.
The application of the radiative wind theory to rotating stars
leads to an expression (cf. Maeder \cite{mae99a}) for the 
mass loss rates $\dot{M} (\vartheta)$ as a function of the 
colatitude $\vartheta$, this expression shows
that there are two main effects influencing the anisotropic mass loss:
~a) the ``$g_{\mathrm{eff}}$''--effect, i.e. the
higher gravity and \te   at the pole of a 
rotating star (due to  von Zeipel's
theorem) enhances the polar mass loss; b) the ``opacity--effect'', i.e. the
higher opacity, and the higher force multipliers, due to the lower temperature
at the equator favour an equatorial ejection and the formation of a ring.
The B[e] stars show both  polar and  equatorial ejections 
(Zickgraf \cite{zick}). The equatorial ejection is strongly favoured by the
so--called bi--stability effect (Lamers \cite{lam}), i.e. a jump of the 
opacity  near 20'000${\degr}$K, i.e. close to spectral type B2 where a maximum 
of the fraction of Be stars is observed.

Models of rotating stars, including hydrostatic distorsion, shear mixing,
meridional circulation, enhanced mass loss rates, loss of angular momentum
etc. have been constructed recently (Meynet \cite{mey99}; 
Maeder \cite{mae99b}). These models show the high influence of rotation on
massive star evolution. 
The main reason is that shear mixing, which is the most efficient mixing
process,
 is favoured by a high thermal diffusivity as observed in massive stars
 (Maeder \cite{mae97}); also,  
other effects such as meridional circulation 
(Maeder \& Zahn \cite{maezah}) 
and anisotropic mass loss are quite significant in massive stars. In 
view of the large consequences of rotation,
it is particularly useful to examine whether the Be fraction is systematically
higher in clusters of lower metallicities, a possibility 
also suspected by Grebel et al.\ (\cite{grericboe}, \cite{gre}) and Mazzali et
al. (\cite{maz}). Section 2 establishes the number data of Be stars in an 
ensemble of clusters of the Galaxy, LMC and SMC. The relation of the results
with metallicity is examined in Section 3, together with abundance indications 
related to rotation. Section 4 gives the conclusions. 

\section{Number counts of Be stars in clusters}

\subsection{Cluster data}

Since the fraction of Be stars is rapidly changing with cluster ages 
(Mermilliod \cite{mer}), we must carefully separate ages and
metallicity effects. We select clusters in the age range of $\log$ age =
7.00 to 7.40 which have their turnoff around the maximum of the Be star
distribution. The comparison of clusters of 
various ages in different galaxies may
lead to confusion between age and metallicity effects. 

For the Milky Way the clusters were selected from the cluster data base WEBDA
(Mermilliod \cite{mer99}). This data base contains homogeneous photometric 
data with consistent estimates of reddening, distance moduli and ages. 
WEBDA also provides indications on known
binaries, stellar peculiarities and Be star identifications based on standard
spectroscopic criteria. The data base includes the recent survey of
stars with H$\alpha$ emission by Kohoutek and Wehmeyer(\cite{koh}).
 Some 11 well studied
clusters of the Galaxy are in the age interval 
considered (Table 1). On the basis of their galactocentric distances,
the clusters are separated in two groups: one outside and one
inside the solar location.

\begin{figure}[htbp]
\centerline{\hbox{\psfig{figure=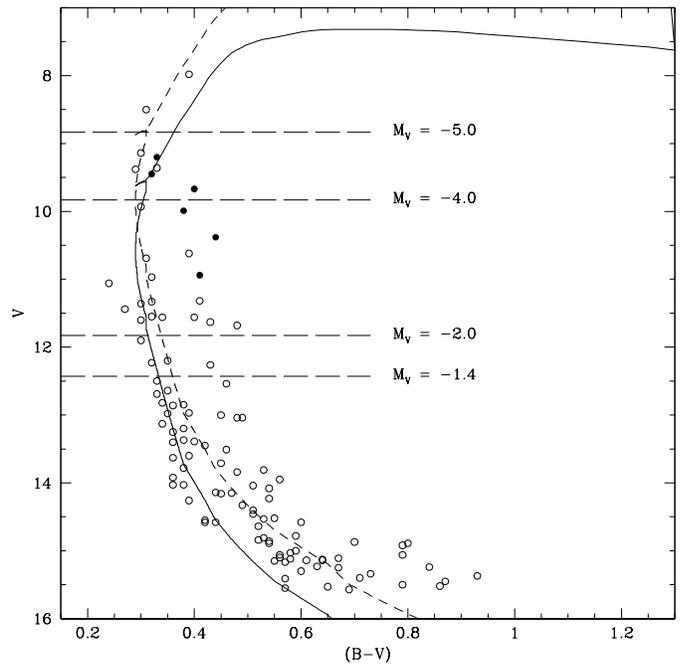,width=\linewidth}}}
\caption{Example of colour--magnitude diagram used for the number counts
of Be and B stars in the case of NGC 884 from the WEBDA data base
(Mermilliod \cite{mer99}). The limits of the various  magnitude intervals 
are shown: $M_{\mathrm{V}}$ = -4, -2; $M_{\mathrm{V}}$ = -5, -2;
$M_{\mathrm{V}}$
 = -5, -1.4. Be stars are represented by black dots.
The isochrone corresponding to log age = 7.1 is represented
by a continuous line, and the top of the binary sequence
upwards shifted by 0.75 mag. is shown by a broken line.}\label{Fig1}
\end{figure} 

 For the LMC the photometric data, the age
determinations and the identifications of Be stars come from Grebel
(\cite{gre}),
Dieball and Grebel (\cite{die}), Keller et al. (\cite{kel2}),
and Grebel and Chu (in preparation). 
 A distance
modulus of 18.50 is taken in agreement with current 
determinations (cf. Cole
\cite{col}). Nine LMC clusters (listed in Table 1) are in the considered age
interval. For the SMC the typical cluster NGC 330 
is analysed with the data from
Grebel et al. (\cite{grerob}). Unfortunately there is no other cluster with
reliable Be star identification in or even 
near the  considered age range. The
cluster NGC~346 of the SMC, which contains a few Be stars, is much too young
and has a turnoff far out of the age range where the maximum of Be stars
 occurs. Kudritzki et al. (\cite{kud}) give an age of $2.6 \cdot 10^6$~yr
for NGC 346, which is consistent with the presence of stars up to about 100
M$_\odot$ (Massey et al. \cite{mas}). This is in agreement with other
authors (Cassatella et al. \cite{cas}; Haser et al.
 \cite{has}) who notice the presence of O3 stars in NGC~346.

\setlength{\tabcolsep}{4mm}
\begin{table*}[htbp]
\caption[]{Cluster data}
\begin{flushleft}
\begin{tabular}{lrrrrrrrrl}
\noalign{\smallskip}
\hline
\noalign{\smallskip}
\multicolumn{1}{r}{} & 
\multicolumn{1}{r}{} & 
\multicolumn{1}{r}{} & 
\multicolumn{2}{c}{$M_{\mathrm{V}}$ = -5,\,-1.4} & 
\multicolumn{2}{c}{$M_{\mathrm{V}}$ = -5,\,-2} &
\multicolumn{2}{c}{$M_{\mathrm{V}}$ = -4,\,-2} & 
\multicolumn{1}{r}{}\\[2mm]
Cluster & $\log$(age)\hspace{-3mm} & DM\hspace{1mm} & B+Be\hspace{-2mm} & Be &
B+Be\hspace{-2mm} & Be & B+Be\hspace{-2mm} & Be & Ref.\\
\noalign{\smallskip}
\hline
\noalign{\smallskip}
{\bf SMC} &&&&&&&&&\\
NGC 330 & 7.30 & 18.90 & 128 & 50 & 89 & 39 & 81 & 37 & Grebel et al.
1996\\[3mm]
{\bf LMC} &&&&&&&&&\\
Hodge 301 & 7.30 & 18.50 & 44 & 10 & 25 & 9 & 25 & 9 & Grebel \& Chu in prep\\
KMHK 1019 & 7.20 & 18.50 & 11 & 2 & 1 & 1 & --- & --- & Dieball \& Grebel
1998\\
NGC 1818 A & 7.40 & 18.50 & 94 & 34 & 58 & 22 & 51 & 19 & Grebel 1997\\
NGC 1818 B & 7.40 & 18.50 & 7 & 3 & 4 & 2 & 4 & 2 & Grebel 1997\\
NGC 1948 & 7.40:\hspace{-1mm} & 18.50 & 101 & 11 & 71 &  9 &
64 &  9 & Keller et al. 1999\\
NGC 2004 & 7.40 & 18.50 & 130:\hspace{-1mm} & 25 & 96:\hspace{-1mm} & 21 & 
82:\hspace{-1mm} & 17 & Grebel 1998\\
NGC 2006 & 7.30 & 18.50 & 35 & 10 & 21 & 5 & 21 & 5 & Dieball \& Grebel 1998\\
NGC 2100 & 7.40:\hspace{-1mm} &
 18.50 & 67 & 19 & 49 & 16 & 43 & 15& Keller et al. 1999\\
SL 538 & 7.20 & 18.50 & 46 & 11 & 29 & 9 & 28 & 9 & Dieball \& Grebel 1998\\
TOTAL &&& 535 & 125 & 354 & 94 & 318 & 85 & \\[3mm]
{\bf Galaxy ext.} && 
{$\mathrm{R_{gc}}(kpc)$}\hspace{-5mm} &&&&&&&\\
NGC 457 & 7.30 & 9.92 & 28 & 4 & 17 & 4 & 13 & 1 & Mermilliod 1999\\
NGC 581 & 7.20 & 9.54 & 8 & 1 & 5 & 0 & 4 & 0 & Mermilliod 1999\\
NGC 663 & 7.20 & 9.61 & 35 & 12 & 23 & 12 & 21 & 10 & Mermilliod 1999\\
NGC 869 & 7.10 & 9.60 &42 & 3 & 33 & 3 & 28 & 2 & Mermilliod 1999\\
NGC 884 & 7.10 & 9.91 & 28 & 6 & 24 & 6 & 18 & 3 & Mermilliod 1999\\
NGC 957 & 7.15 & 9.41 & 15 & 4 & 13 & 4 & 12 & 3 & Mermilliod 1999\\
NGC 2439 & 7.30 & 10.45 & 31 & 5 & 21 & 5 & 21 & 5 & Mermilliod 1999\\
NGC 7160 & 7.20 & 8.22 & 3 & 1 & 2 & 1 & 2 & 1 & Mermilliod 1999\\
TOTAL &&& 190 & 36 & 138 & 35 & 119 & 25 &\\[3mm]
{\bf Galaxy int.} &&&&&&&&&\\
NGC 3293 & 7.20 & 7.70 & 37 & 1 & 32 & 1 & 25 & 1 & Mermilliod 1999\\
NGC 3766 & 7.40 & 7.44 & 42 & 10 & 25 & 8 & 24 & 8 & Mermilliod 1999\\
NGC 4755 & 7.20 & 7.07 & 47 & 3 & 27 & 3 & 23 & 3 & Mermilliod 1999\\
TOTAL &&& 126 & 14 & 84 & 12 & 72 & 12 &\\
\hline
\end{tabular}
\end{flushleft}
\end{table*}

\subsection{Number counts of Be stars in luminosity intervals}

We choose to proceed to number counts in given intervals of magnitude 
$M_{\mathrm{V}}$ rather than in given ranges of spectral type and \te since
spectral types are not available for all clusters. Also, we noticed that
high rotation strongly modifies the average \te and very little the average
stellar luminosities (Maeder and Peytremann \cite{maepey}). Furthermore,
massive stars of the same mass, but different metallicites, have rather large
differences in \te, while this is 
 not the case for the luminosities (Schaller et al. \cite{scha}).
In order to test the independence of the results with respect to the 
chosen $M_{\mathrm{V}}$ range we have done number counts in three 
intervals of $M_{\mathrm{V}}$ centered on the domain O9 to B3, i.e. 
$M_{\mathrm{V}}$ = -4, -2; $M_{\mathrm{V}}$ = -5, -2; $M_{\mathrm{V}}$ =
-5, -1.4 (the limit $M_{\mathrm{V}}$ = -1.4 corresponds to the spectral type
B3 according to the calibration of Zorec and Briot (1991) for main sequence
B--type stars). Fig. \ref{Fig1} shows an example of 
the HR diagram used for the number counts in the 
case of NGC 884 ($\chi$ Persei cluster) with the various 
magnitude intervals considered.

Table 1 shows the number counts for
the 19 clusters in the four different locations
considered. The 2nd column gives the $\log$ of the age, the 3rd column gives
the distance moduli in the SMC and LMC, the galactocentric distance for 
galactic clusters is given in the 3rd column as well; the following columns
give the numbers of B plus Be stars, and those 
of Be stars only in the indicated
magnitude intervals. The reference for the data is in the last column.\\
The average number ratios $Be/(B+Be)$ stars in the four zones
and for the different magnitude intervals considered are given in Table 2.
We clearly notice a systematic trend for smaller $Be/(B+Be)$ 
number ratios in the
sequence of the four groups considered from SMC to LMC to the galactic
exterior and interior. The total difference in the $Be/(B+Be)$ ratios
amounts to  a factor of about three, which is important and also
systematic in relation with the metallicities as discussed below.

\subsection{Possible relations with Z}

We examine the possible relation between
the fraction of Be stars and the
local metallicity Z in the regions where these stars were formed. 
Let us notice that the 
suggestion to determine whether the frequency of Be stars is somehow
related to the metal abundance was also made by Grebel et al.
(1992) and Mazzali et al. (1996).

\begin{table}[htbp]
\caption[]{Number ratios of Be to B+Be stars in the Magellanic Clouds
and Milky Way.}
\begin{flushleft}
\begin{tabular}{lllll}
\noalign{\smallskip}
\hline
\noalign{\smallskip}
\multicolumn{1}{l}{} & 
\multicolumn{1}{l}{$M_{\mathrm{V}}$} &
\multicolumn{3}{c}{$\frac{Be}{B+Be}$}\\
\noalign{\smallskip}
\cline{3-5}
\noalign{\smallskip}
\multicolumn{1}{l}{} & 
\multicolumn{1}{l}{} & 
\multicolumn{1}{l}{-5,\,-1.4} &
\multicolumn{1}{l}{-5,\,-2} & 
\multicolumn{1}{l}{-4,\,-2}\\ 
\noalign{\smallskip}
\hline
\noalign{\smallskip}
SMC & 128 & .39 & .44 & .46\\
LMC & 535 & .23 & .27 & .27\\
Galaxy ext. & 190 & .19 & .25 & .21\\
Galaxy int. & 126 & .11 & .14 & .17\\
\noalign{\smallskip}
\hline
\noalign{\smallskip}
\end{tabular}
\end{flushleft}
\end{table}

We use  the  recent data on the metallicity Z for the
various zones. For the solar metallicity we take a value of Z$_\odot$ =
0.018, which is consistent with recent solar models and heliosismological
data (Brun et al. 1998). For the chemical gradient in the Galaxy the
current values range between $\Delta[O/H]/kpc = -0.07$
(cf. Shaver et al. \cite{sha}; Gummersbach et al. \cite{gum}) and -0.05 (cf. 
Vilchez and Esteban \cite{vil}). The group of clusters towards
the anticenter has an average galactocentric distance $R_{gc}$ of about 1.5
kpc larger than that of
the Sun (taken to be $R_{gc} = 8$ kpc), while this
average difference is 0.6 kpc for the group of clusters towards the
galactic interior. Thus we consider that the
 average local metallicities for the exterior and
interior groups are respectively Z\,=\,0.014 and Z\,=\,0.020  for a
galactic gradient of -0.07. For a gradient of -0.05 the Z-values would
be 0.015 and 0.019 respectively, i.e. not very different 
from the latter values.

For the young population in the LMC the range of estimated metallicities 
is rather large. For example, Olszewski et al. (\cite{ols}) give [Fe/H] 
values in the range
-0.3 to -\,0.42, which corresponds to Z\,=\,0.010 to 0.008.   
Jasniewicz and
Th\'evenin (1994) find values from [Fe/H] = -0.4 for NGC 1818 to -0.55 for
NGC 2004, which correspond to respectively Z\,=\,0.008 and 0.006. Bica et
al. (\cite{bic}) obtained values  Z\,=\,0.004 to 0.005. Luck et al.
(\cite{luck}) find a mean [Fe/H] = $-0.34\pm0.15$ for LMC field Cepheids. 
In this context, an average value of 0.007 seems appropriate. For the SMC
field Reiterman et al. (\cite{rei}), Spite et al. (\cite{spi}), Grebel and
Richtler (\cite{greric}) give values corresponding to Z\,=\,0.001 to
0.003.  
For NGC 330, Hilker et al. (\cite{hil}) give
a value 
Z\,=\,0.002, and the results of Hill (\cite{hill}) correspond to 0.003
to 0.004 and those of Oliva et al. (\cite{oli}) to 0.001.
Luck et al. (\cite{luck}) find a value corresponding to Z=0.004.
  Gonzalez and Wallerstein (1999 \cite{gon}) find Z = 0.002.  In this
context an average value Z\,=\,0.002 for the SMC is appropriate. 

Fig. 2 shows the relation between the fraction $Be/(B+Be)$ and the local
average metallicity. The trend is quite clear for the various magnitude 
intervals considered. For the sample of the 4 locations there seems to be
a clear increase of the fraction of the relative number of Be stars with
the local average initial metallicity.

\begin{figure}[htbp]
\centerline{\hbox{\psfig{figure=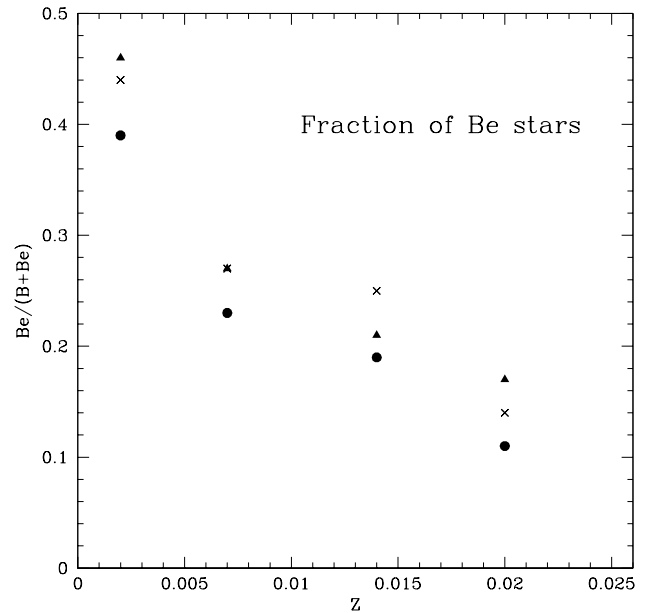,width=\linewidth}}}
\caption{Relation between the number ratio Be/(B+Be) and the
local metallicity for the 4 groups of clusters considered in Table 
2. To test the validity of the results,
the  number counts were made in different magnitude intervals,
the dots refer to counts made in the magnitude interval
$M_{\mathrm{V}}$ = -5, -1.4, the crosses to the 
interval -5, -2 and the
triangles to the interval -4, -2.}\label{Fig2}
\end{figure} 

\section{Interpretation of the results}

There are various possibilities to interpret the above results.

\noindent
\parbox[t]{.6cm}{1.}\parbox[t]{8cm}{Statistical effects related to the 
intermittence of the Be phenomenon.}\\[2mm]
\parbox[t]{.6cm}{2.}\parbox[t]{8cm}{The visibility of the Be phenomenon
is enhanced at low metallicities.}\\[2mm]
\parbox[t]{.6cm}{3.}\parbox[t]{8cm}{Shell ejection is favoured at low
metallicities.}\\[2mm]
\parbox[t]{.6cm}{4.}\parbox[t]{8cm}{Differences in the distribution
of the rotational velocities.}\\

The first possibility is very unlikely. 
In the Milky Way Be stars have generally 
been observed over a longer period than in the LMC or SMC. Therefore,
the intermittent Be star have a higher chance to be detected in
the Milky Way. Consequently, the correction of this possible effect would
increase the trend observed in Fig. 2 rather than reduce it. Indeed, in
the SMC cluster NGC 330 the Be phenomenon has been found to be 
intermittent in a number of stars re-observed over the period of several
years (Grebel 1995, Keller et al. 1999).  The second
hypothesis also seems difficult to sustain, because at higher metallicities
the ejected shell and circumstellar material would be more opaque and contain
more dust, thus their visibility  through emission lines as well as by
their infrared continuum would be enhanced. Thus, the correction
of such an effect (if feasible) would increase the observed trend.

The same kind of remark holds for the third hypothesis. Models of rotating
stars with mass loss (Maeder \cite{mae99a}) show two effects: a) polar
ejection is favoured in rotating stars by the higher \te at the pole;
b) equatorial ejection of a shell is favoured by larger opacities.
Thus, at lower Z, it is not expected that shell ejection is favoured
in the equatorial regions of a rotating star,
because the opacities are lower.
 The possible effect of metallicity on
the terminal velocities, which may follow from the wind-compressed disk
model (Bjorkman and Cassinelli \cite{bjo}; Grebel \cite{gre}), is
found to be negligible (Maeder \cite{mae99a}).

Thus we are left with the possibility that the higher 
fraction of 
Be stars at lower Z  is the signature of more fast rotators at lower Z. 
Obviously, direct measurements in nearby galaxies
 are very needed and envisaged in order to confirm or
reject this suggestion.  For NGC 330
measurements of projected rotational velocities
($v\,\sin\,i$) by Mazzali et al. (\cite{maz}) and Keller and Bessell 
(\cite{kel1}) show values of up to 400 km s$^{-1}$.  
Furthermore, a recent study 
of nitrogen abundances done by Venn (\cite{ven}) offers a strong support 
to the possibility of faster rotation in NGC 330 of the SMC. 

Indeed, several authors have found that B-- and A--type
supergiants in the Galaxy, LMC and SMC show nitrogen enhancements not
predicted by standard evolutionary models (cf. Lennon et al. \cite{len};
Fitzpatrick and Bohannan \cite{fit}; Venn \cite{ven}, \cite{vencar}).
The current interpretation of these surface enhancements requires some
additional mixing processes, possibly rotational mixing (cf. Langer
\cite{lan}; Meynet \cite{mey}; Maeder
and Zahn \cite{maezah}). New rotating models
show consistently some N--enrichments already present for medium rotation
from the end of the MS onwards (Meynet \cite{mey}). A remarkable point
is that part of the N--enhancement observable in B-- and A--type supergiants
could be of primary origin,
 due to the fact that some new $^{12}$C resulting from
the 3$\alpha$ burning is rotationally diffused into the H--burning shell,
where the CNO cycle converts this ``new $^{12}$C'' into $^{14}$N, which
is thus of primary origin. 

Another interesting fact recently found by Venn (\cite{vencar}) is that the
degree of N enhancement for A supergiants is much higher in the SMC than in
the Milky Way. For galactic supergiants the typical enhancement of [N/H]
is about a factor of 2, while in the SMC it reaches a factor of 10--20
with respect to the standard local average [N/H] in the SMC. 
As a matter of fact, converting the $^{12}$C entirely into $^{14}$N by
the end of the CN cycle (which is dominant) would lead to an increase of
[N/H] by a factor of 4--5, but not as high as 20. Thus the
high [N/H] enhancement observed for SMC supergiants might be a sign of
primary nitrogen. We do not know, the only way to clarify
is to measure the sum of CNO elements and check whether it is the same
or not in supergiants as in MS stars. Anyhow, whether this is primary
or secondary nitrogen, the point for now is that
this higher [N/H] is the signature of much more
mixing in the SMC than in the Galaxy, a fact 
 which is quite consistent with
the above suggestion of faster rotation for massive stars in the SMC. 
 The possibility of faster rotation in the SMC has also
been mentioned by Venn (\cite{vencar}) and the results presently
available on the Be star fractions as a function of Z clearly support this
view. 

\hspace{0.1mm}

\section{Conclusions}

The main result is that the relative fraction of Be stars 
with respect to all B stars in the spectral interval O9 to B3
is increasing for lower metallicities Z. This fact together with
the much larger N-excesses observed in A-type supergiants of the
SMC than in galactic supergiants strongly supports the 
suggestion that there are more fast rotators among massive stars at 
lower Z. Of course, direct observations of large numbers of 
$v\hspace{0.3mm}\sin{\hspace{-0.5mm}i}$ in LMC and SMC clusters are
very much needed in order to further substantiate the above results. 

We may wonder about the origin of the possible
 higher rotation velocities at low
metallicities. This origin is likely related to some metallicity
effects in the process of star formation. There are many possibilities,
 for example we may notice that a lower Z implies 
less dust and ions in star forming regions. The coupling
of the magnetic field to the matter is then weaker, the
ambipolar diffusion of the magnetic field  should
proceed faster, thus leading to less angular momentum losses by 
the central contracting body. Another possibility is that during
pre-main sequence evolution, the weaker opacities at lower Z
are leading to an earlier disappearance of the external convective
zones and thus to less magnetic coupling between the forming
star and its surroundings. Also the importance and the 
survival lifetime of the accretion disk may increase
with metallicity, thus favouring the dissipation of
angular momentum at higher Z. Of
course, numerical models are needed to examine these various tentative
suggestions.

Looking ahead we may also mention that the consequence of faster 
rotation at lower Z may be considerable, with large differences in
evolutionary tracks, lifetimes and nucleosynthesis. In particular
we know that current models are unable to account for the occurrence
of the large quantity of red supergiants in the SMC for which more
mixing would be needed (cf. Langer and Maeder \cite{lan}). In the
context of nucleosynthesis the possible 
formation of primary nitrogen at low 
Z is interesting in relation with the suggestion that primary 
nitrogen is needed in the early chemical evolution of galaxies
(cf. Matteucci and Fran\c{c}ois \cite{mat}; Pagel \cite{pag}).
Appropriate stellar models would bring new insight into the interpretation
of star populations in low Z galaxies, like blue compact
galaxies or galaxies at cosmological distances like
in the Hubble Deep Field. 

\begin{acknowledgements}
We thank Dr. Kohoutek for having provided a printed and an e-version 
of his catalogue to one of us (J.-C. M.).  EKG gratefully 
acknowledges support by NASA through grant HF-01108.01-98A from the 
Space Telescope Science
Institute, which is operated by the Association of Universities for Research
in Astronomy, Inc., under NASA contract NAS5-26555. AM expresses his
thanks to Dr. Hans Zinnecker for nice discussions on the 
possible origin of faster rotation at lower metallicities.
\end{acknowledgements}

\end{document}